\documentstyle[12pt]{article}
\begin{document}
\begin{center}
\vspace{2.0cm}
{\large {\bf The generation model of particle physics and the cosmological matter-antimatter asymmetry problem}}\\
\vspace{0.5cm}
{\small B.A. Robson$^{*}$ \\
{\small \it Department of Theoretical Physics, Research School of Physics and Engineering,}\\
{\small \it The Australian National University, Acton ACT 2601, Australia} \\
{\small \it $^{*}$brian.robson@anu.edu.au}} \\
\date{\small (received...............................) }
\end{center}

\vspace{1.0cm}

\noindent The matter-antimatter asymmetry problem, corresponding to the virtual nonexistence of antimatter in the universe, is one of the greatest mysteries of cosmology. Within the framework of the Generation Model (GM) of particle physics, it is demonstrated that the matter-antimatter asymmetry problem may be understood in terms of the composite leptons and quarks of the GM. It is concluded that there is essentially no matter-antimatter asymmetry in the present universe and that the observed hydrogen-antihydrogen asymmetry may be understood in terms of statistical fluctuations associated with the complex many-body processes involved in the formation of either a hydrogen atom or an antihydrogen atom.\\


\noindent PACS Number(s): 12.60.Rc, 95.30.Cq, 98.80.Bp \\

\newpage

\noindent {\bf 1. Introduction}\\

\noindent The matter-antimatter asymmetry problem, corresponding to the virtual nonexistence of antimatter in the universe, is one of the greatest mysteries of cosmology, along with dark matter and dark energy.$^{1}$ Recently, an understanding of both dark matter and dark energy has been provided by the development of a new quantum theory of gravity,$^{2,3}$ which is based on the Generation Model (GM) of particle physics.$^{4}$ In this paper it is demonstrated that the matter-antimatter asymmetry problem may also be understood in terms of the composite leptons and quarks of the GM, contrary to the elementary leptons and quarks of the Standard Model (SM) of particle physics.$^{5}$\\

According to the SM of particle physics, the universe is made essentially of matter comprising three kinds of elementary particles: electrons, up quarks and down quarks. An electron has electric charge $Q = -1$ but has no color charge. The up and down quarks have electric charges $Q = +\frac{2}{3}$ and $Q = -\frac{1}{3}$, respectively, and have a single color charge, red, green or blue.\\

Each of the above elementary particles has an elementary antiparticle, which has the opposite electric and color charges: the antielectron (positron) has electric charge $Q = +1$ and has no color charge, the antiup and antidown quarks have electric charges $Q = -\frac{2}{3}$ and $Q = +\frac{1}{3}$, respectively, and have a single color charge, antired, antigreen or antiblue.\\

Experiment indicates that an antiparticle may annihilate {\bf only} with its corresponding particle producing ``interaction mediating particles", such as photons, gluons or weak W/Z bosons, while conserving both energy and momentum. For example, an electron and a positron may annihilate each other producing photons or at sufficiently high energy, W/Z bosons. On the other hand an electron and an up quark do not annihilate. An up quark carrying a red color charge may annihilate with an antiup quark carrying an antired color charge to produce photons. This particle-antiparticle pair may also annihilate to produce massless electrically neutral gluons carrying both a red color charge and an antired color charge thereby conserving color charge. Gluons act as the exchange particles for the strong interaction between color charged quarks and are analogous to the exchange of photons in the electromagnetic interaction between two electrically charged particles.\\

However, the nature of the gluon fields of the strong interaction and that of the photon fields of the electromagnetic interaction are quite different. In particular, gluons carry color charges and self-interact, while photons, being electrically neutral, do not. This property of gluons leads to antiscreening effects$^{6,7}$ of the strong interaction, contrasting with the screening effects of the electromagnetic interaction.\\

The nature of the gluon fields is such that they lead to a runaway growth of the fields surrounding an isolated color charge.$^{8}$ In fact all this structure, via virtual color charged gluons, implies that an isolated quark would have an infinite energy associated with it. This is the reason why isolated quarks are not seen. Nature requires these infinities to be essentially cancelled or at least made finite. It does this for hadrons in two ways: either by bringing a quark and an antiquark close together (i.e., forming a meson) or by bringing three quarks, one of each color, together (i.e., forming a baryon) so that in each case the composite hadron is colorless (color neutral). Thus the colored quarks of the SM form composite colorless baryons, protons and neutrons, which presently comprise most of the matter in the universe.\\

According to the prevailing cosmological model,$^{1}$ the universe was created in the so-called `Big Bang' from pure energy, and is currently composed of about 5\% ordinary matter, 27\% dark matter and 68\% dark energy. The matter described by the SM refers to the 5\% ordinary matter, which prior to the nucleosynthesis, i.e., the fusion into heavier elements, consisted of about 92\% hydrogen atoms and 8\% helium atoms,$^{9}$ so that the ordinary matter of the universe was, and still is, essentially electrically neutral and colorless.\\

It is generally considered that the Big Bang and its aftermath produced equal numbers of particles and antiparticles, although the universe today appears to consist almost entirely of matter rather than antimatter. This constitutes the matter-antimatter asymmetry problem: where have all the antiparticles gone? Currently there is no acceptable understanding of this problem.\\

Since the physical nature of the Big Bang is still not understood, it is not possible to discuss the matter-antimatter asymmetry problem from the initial singularity. Consequently, the matter-antimatter asymmetry problem will be discussed in terms of the observed nature of the universe.\\

For many decades now the SM has been unable to provide an acceptable understanding of the matter-antimatter asymmetry problem. In Sec.\ 2 the main attempts, within the framework of the SM, to understand the asymmetry problem will be discussed briefly. In Sec.\ 3 we shall introduce the GM and summarize the essential differences between the GM and the SM. In Sec.\ 4 we shall show how the GM naturally provides an understanding of the asymmetry problem in terms of composite leptons and quarks. Section 5 states the conclusions.\\

\vspace{0.5cm}

\noindent {\bf 2. Matter-Antimatter Asymmetry Problem and the Standard Model}\\

\noindent Within the framework of the SM the matter-antimatter asymmetry problem is generally considered to be related to the baryon asymmetry problem, i.e., the imbalance of baryonic matter and antibaryonic matter in the observable universe. The universe seems to consist almost entirely of hydrogen and helium atoms rather than antihydrogen and antihelium atoms.\\

This observation is at first sight rather surprising since the origin of the universe in the Big Bang is generally considered to have produced equal numbers of baryons and antibaryons, which as the universe cooled should have annihilated in pairs back to pure energy, so that the universe would be empty of matter. In the SM this is assumed to imply that although the universe was originally perfectly symmetric in baryons and antibaryons, during the cooling period some physical processes contributed to a small imbalance in favor of baryons.\\
 
The small excess of baryons over antibaryons indicates that contrary to the current laws of physics, baryon number must be violated in some physical process. This was proposed by Sakharov$^{10}$ in 1967. Sakharov proposed a set of three necessary conditions, within the framework of the SM, that a physical process must satisfy to produce baryons and antibaryons at different rates: (i) violation of baryon number; (ii) violation of both charge conjugation symmetry, C, and charge conjugation-parity symmetry, CP; and (iii) the process must not be in thermal equilibrium. Violation of baryon number is required to produce an excess of baryons over antibaryons, while C symmetry violation ensures the non-existence of processes, which produce an equivalent excess of antibaryons over baryons. Similarly, violation of CP symmetry is required so that equal numbers of left-handed baryons and right-handed antibaryons as well as equal numbers of right-handed baryons and left-handed antibaryons are not produced. Finally, a departure from thermal equilibrium must play a role so that CPT symmetry does not ensure compensation between processes increasing and decreasing the baryon number.$^{11}$\\

The first Sakharov criterion: violation of baryon number, would be achieved if antiprotons or protons decayed into lighter subatomic particles such as a neutral pion and an electron or positron, respectively. However, there is currently no experimental evidence that such `direct' violations of baryon number occur.\\

Thus researchers turned their attention to `indirect' violations of baryon number, which are concerned with Sakharov's second criterion: CP violation, which indicates the possibility that some physical processes may distinguish between matter and antimatter.\\

Both the electromagnetic and strong interactions are symmetric under C and P, and consequently they are also symmetric under the product CP. However, this is not necessarily the case for the weak interaction, which violates both C and P symmetries. Indeed, the 1964 discovery$^{12}$ of the decay of the long-lived $K^{0}$ meson to two charged pions brought about the surprising conclusion that CP is also violated in the weak interaction. The violation of CP in weak interactions implies that such physical processes could lead to indirect violation of baryon number so that matter creation would be preferred over antimatter creation.\\

In the SM CP violation originates from charge-changing (CC) weak interactions that change the charge and flavor of quarks. The six known quarks consist of three up-like quarks with charge $Q = +\frac{2}{3}$: up ($u$), charmed ($c$) and top ($t$); and three down-like quarks with charge $Q = -\frac{1}{3}$: down ($d$), strange ($s$) and bottom ($b$). The CC weak interactions cause each up-like quark to turn into a down-like quark and vice-versa. The transition amplitudes for the nine combinations are given by the Cabibbo-Kobayashi-Maskawa (CKM) matrix elements.$^{13,14}$\\

Prior to the quark model of Gell-Mann$^{15}$ and Zweig$^{16}$, Cabibbo$^{13}$ in 1963 had introduced some of the matrix elements of the CKM matrix in order to preserve the universality of the CC weak interaction. In 1973 Kobayashi and Maskawa$^{14}$ discovered that for the case of three generations of quarks the CC weak interactions may violate CP. However, the Kobayashi-Maskawa CP violation was found to be tiny, primarily because of the smallness of the relevant matrix elements. Consequently, any physical process that produces more matter than antimatter would have been ineffectual. Although the excess of matter over antimatter is generally considered to have been only one part in a billion, the effect of the Kobayashi-Maskawa CP violation process falls far short of even this very small amount by many orders of magnitude. Indeed it is estimated that the baryon excess produced by the Kobayashi-Maskawa CP violation process is only sufficient to provide the baryons of a single galaxy in the universe, which comprises billions of galaxies.\\

The third criterion of Sakharov: departure from thermal equilibrium, is generally assumed to also occur within the electroweak sector of the SM during the so-called electroweak phase transition.$^{17}$ This is assumed to be a first order transition between the state in which the $W$ and $Z$ gauge bosons are massless to a state in which they are massive. The massive $W$ and $Z$ bosons are assumed to arise as a result of some unknown mechanism which breaks the electroweak symmetry, and it is during this electroweak symmetry breaking that departure of thermal equilibrium takes place.\\

Thus the SM does provide possible physical processes, which satisfy all three necessary criteria of Sakharov. However, the assumed physical processes do not seem capable of providing an acceptable explanation for the matter-antimatter asymmetry. The general conclusion is that physics beyond the SM is required for this purpose.\\

\vspace{0.5cm}

\noindent {\bf 3. The Generation Model }\\

\noindent The Generation Model (GM) has been developed over many years$^{18-22}$ and is an alternative model to the SM of particle physics. The current version of the GM is described in detail in Ref.\ 4. There are three essential differences between this GM and the SM: (i) the classification of the leptons and quarks in terms of additive quantum numbers; (ii) the roles played by the mass eigenstate quarks and the weak eigenstate quarks and (iii) the nature of the weak interactions. These are discussed in detail in Ref.\ 23. For the purpose of the present paper, only the essential differences (i) and (ii) will be discussed more briefly below.\\ 

The GM employs a unified classification scheme for the leptons and quarks rather than the non-unified scheme of the SM. This unified system is based upon the use of only three additive quantum numbers rather than the nine independent additive quantum numbers of the SM. These three additive quantum numbers are charge ($Q$), particle number ($p$) and generation quantum number ($g$). The non-unified system of the SM uses four additive quantum numbers for leptons: charge ($Q$), lepton number ($L$), muon lepton number ($L_{\mu}$) and tau lepton number ($L_{\tau}$) plus six additive quantum numbers for quarks: charge ($Q$), baryon number ($A$), strangeness ($S$), charm ($C$), bottomness ($B$) and topness ($T$).\\

Comparison of the two models, SM and GM, indicates that they have only one additive quantum number in common, namely electric charge $Q$. The second additive quantum number of the GM, particle number $p$ replaces both lepton number $L$ and baryon number $A$ of the SM. The third additive quantum number of the GM, generation quantum number $g$ effectively replaces the remaining six additive quantum numbers of the SM. Thus the GM provides both a simpler and {\it unified} classification scheme for leptons and quarks. Indeed the GM classification scheme indicates that leptons and quarks are intimately related and led to composite models of leptons and quarks, which in turn led to new paradigms for both mass and gravity.$^{22,24}$\\

An important feature of the GM classification scheme is that all three additive quantum numbers, $Q$, $p$ and $g$, are required to be conserved in all leptonic and hadronic processes. In particular, the generation quantum number $g$ is strictly conserved in weak interactions unlike some of the quantum numbers of the SM, e.g., strangeness $S$. This latter requirement led to a new treatment of quark mixing in hadronic processes$^{21}$: the GM differs from the SM in two fundamental ways, which are essential to preserve the universality of the CC weak interaction for hadronic processes.\\
 
First, the GM postulates that the mass eigenstate quarks of the same generation, e.g., ($u$, $d$), form weak isospin doublets and couple with the full strength of the CC weak interaction, like the lepton doublets. Unlike the SM, the GM requires that there is no coupling between mass eigenstate quarks from different generations. This latter requirement corresponds to the conservation of the generation quantum number $g$ in the CC weak interaction processes.\\

Second, the GM postulates that hadrons are composed of weak eigenstate quarks, rather than the corresponding mass eigenstate quarks as in the SM. Thus the GM differs from the SM in that it treats quark mixing differently from the method introduced by Cabibbo$^{13}$ and employed in the SM. Essentially, in the GM, the quark mixing is placed in the quark states (wave functions) rather than in the CC weak interactions. Thus in the GM, the proton is considered to consist of two up quarks and one mixed ($d'$) quark, which is a linear superposition of the down quark, the strange quark and the bottom quark:
\begin{eqnarray}
d' = V_{ud}d + V_{us}s + V_{ub}b~,
\end{eqnarray}
where $V_{ij}$ are the elements of the CKM matrix,$^{13,14}$ rather than two up quarks and one down quarks as in the SM. This allows a unified and simpler classification of both leptons and quarks in terms of only three additive quantum numbers, $Q$, $p$ and $g$, each of which is conserved in all interactions. It should be noted that in both the GM and the SM, the weak eigenstate up quark is assumed to be identical with the mass eigenstate up quark, i.e., there is no mixing with the other up-like quarks, ($c$) and ($t$), unlike the down-like quarks.\\ 

The development of the unified GM classification scheme for leptons and quarks indicated that leptons and quarks are intimately related and led to the development of composite versions of the GM. It should be noted that this is not possible in terms of the non-unified classification scheme of the SM, involving different additive quantum numbers for leptons than for quarks and the non-conservation of some additive quantum numbers, such as strangeness, in the case of quarks.\\

Thus another essential difference between the GM and the SM is that in the GM, the leptons and quarks are composite particles rather than elementary particles as in the SM. In the GM, both leptons and quarks have a substructure, consisting of spin-1/2 massless particles, rishons and/or antirishons, each of which carries a single color charge. The constituents of leptons and quarks are bound together by strong color interactions, mediated by massless hypergluons, acting between the color charged rishons and/or antirishons. These strong color interactions of the GM are analogous to the strong color interactions of the SM, mediated by massless vector gluons, acting between color charged elementary quarks and/or antiquarks. In the GM, the strong color interaction has been taken down one layer of complexity to describe the composite nature of leptons and quarks. The only essential difference between the strong color interactions of the GM and the SM is that the former acts between color charged rishons and/or antirishons, while the latter acts between color charged elementary quark and/or antiquarks. For historical reasons we use the term ``hypergluons" for the mediators of the strong color interactions at the rishon level, rather than the term ``gluons" as employed in the SM, although the effective color interaction between composite quarks and/or composite antiquarks is very similar to that between the elementary quarks and/or elementary antiquarks of the SM. The substructure of leptons and quarks is described in detail in Ref.\ 4.\\

In the GM the elementary rishons and antirishons are required to have the same three kinds of additive quantum numbers as the composite leptons and quarks. Table 1 gives the three additive quantum numbers allotted to the three kinds of rishons of the GM, $T$, $V$ and $U$. For each rishon additive quantum number $N$, the corresponding antirishon has the additive quantum number $-N$. In particular, each rishon has $p = +\frac{1}{3}$, while each antirishon has $p = -\frac{1}{3}$. Thus the particle number $p$ allotted to a composite lepton or quark reflects its degree of particle or antiparticle nature.\\

\begin{center}
\begin{tabular}{|lccc|} \hline
 rishon &~~$Q$~~ &~~$p$~~ &~~$g$~~ \\ \hline
~$T$&$+\frac{1}{3}$&$+\frac{1}{3}$&~~$0$~~ \\
~$V$&$0$&$+\frac{1}{3}$&~~$0$~~ \\
~$U$&$0$&$+\frac{1}{3}$&$-1$~~ \\
\hline
\end{tabular}
\end{center}
\hspace*{3.0cm} Table 1. GM additive quantum numbers for rishons\\

Table 2 displays both the structures and their additive quantum numbers of the first generation of composite leptons and quarks in the GM.\\

\begin{small}
\begin{center}
\begin{tabular}{|lcccc|} \hline
 particle &~~$structure$~~ &~~$Q$~~ &~~$p$~~ &~~$g$~~ \\ \hline
~$e^{+}$&$TTT$&$+1$&$+1$&~~$0$~~ \\
~$u$&$TT\bar{V}$&$+\frac{2}{3}$&$+\frac{1}{3}$&~~$0$~~ \\
~$\bar{d}$&$T\bar{V}\bar{V}$&$+\frac{1}{3}$&$-\frac{1}{3}$&~~$0$~~ \\
~$\nu_{e}$&$\bar{V}\bar{V}\bar{V}$&$0$&$-1$&~~$0$~~ \\
~$\bar{\nu_{e}}$&$VVV$&$0$&$+1$&~~$0$~~ \\
~$d$&$\bar{T}VV$&$-\frac{1}{3}$&$+\frac{1}{3}$&~~$0$~~ \\
~$\bar{u}$&$\bar{T}\bar{T}V$&$-\frac{2}{3}$&$-\frac{1}{3}$&~~$0$~~ \\
~$e^{-}$&$\bar{T}\bar{T}\bar{T}$&$-1$&$-1$&~~$0$~~ \\
\hline
\end{tabular}
\end{center}
\end{small}
\hspace*{3.0cm} Table 2. GM of first generation of leptons and quarks\\

Each lepton of the first generation is assumed to be colorless, consisting of three rishons (or antirishons), each with a different color (or anticolor), analogous to the baryons (or antibaryons) of the SM. The leptons are built out of $T$-rishons and $V$-rishons or their antiparticles $\bar{T}$ and $\bar{V}$, all of which have generation quantum number $g$ = 0.\\

In the GM it is assumed that each quark of the first generation is a composite of a colored rishon and a colorless rishon-antirishon pair, ($T\bar{V}$) or ($\bar{T}V$), so that the quarks carry a color charge. Similarly, the antiquarks are a composite of an anticolored antirishon and a colorless rishon-antirishon pair, so that the antiquarks carry an anticolor charge.\\

The rishon structures of the second generation particles are assumed to be the same as the corresponding particles of the first generation plus the addition of a colorless rishon-antirishon pair, $\Pi$, where
\begin{eqnarray}
\Pi = [(\bar{U}V) + (\bar{V}U)]/\sqrt{2}~,
\end{eqnarray}
which is a quantum mechanical mixture of ($\bar{U}V$) and ($\bar{V}U$), which have $Q = p = 0$ but $g = \pm{1}$. respectively. In this way, the pattern for the first generation is repeated for the second generation. Eq.\ (2) indicates that the generation quantum number $g$ for each second generation particle has two possible values, $\pm{1}$, although in any given transition the generation quantum is required to be conserved.\\

Similarly, the rishon structures of the third generation particles are assumed to be the same as the corresponding particles of the first generation plus the addition of two $\Pi$ rishon-antirishon pairs so that the pattern of the first and second generation is continued for the third generation. The structure
\begin{eqnarray}
\Pi\Pi = [(\bar{U}V)(\bar{U}V) + (\bar{U}V)(\bar{V}U) + (\bar{V}U)(\bar{U}V) + (\bar{V}U)(\bar{V}U)]/2
\end{eqnarray}
indicates that the generation quantum number for each third generation particle has three possible values $g$ = $0$, $\pm{2}$, although in any given transition the generation quantum number is required to be conserved.\\

Eqs.\ (2) and (3) indicate that the weak eigenstate quark $d'$ has charge $Q = -\frac{1}{3}$ and particle number $p = +\frac{1}{3}$ although each component has different values of the generation quantum number $g$.\\

\vspace{0.5cm}

\noindent {\bf 4. Matter-Antimatter Asymmetry Problem and the Generation Model}\\

\noindent The solution of the matter-antimatter asymmetry problem involves the particle number additive quantum number $p$ of the GM. In particular the values of $p$ corresponding to a weak eigenstate up quark ($p = +\frac{1}{3}$), a weak eigenstate down quark ($p = +\frac{1}{3}$) and an electron ($p = -1$). The values of $p = +\frac{1}{3}$ of the quarks, correspond to the values of their baryon number in the SM, while the value of $p = -1$ of the electron, corresponds to minus the value of the lepton number of the electron in the SM. In the GM, the electron consists entirely of antirishons, i.e., antiparticles, while in the SM it is assumed to be a particle, although there is no a priori reason for this assumption.\\

In the GM the proton is assumed to consist of three weak eigenstate quarks, two up quarks and one down quark, so that the proton has particle number $p = +1$. Consequently, a hydrogen atom, consisting of one proton and one electron has particle number $p = 0$: the hydrogen atom in the GM consists basically of an equal number of rishons and antirishons, so that there is no asymmetry of matter and antimatter there.\\

In the GM the neutron consists of three weak eigenstate quarks, one up quark and two down quarks, so that the neutron also has particle number $p = +1$. Consequently, a helium atom, consisting of two protons, two neutrons and two electrons has particle number $p = +2$: the helium atom in the GM consists of six more rishons than antirishons, i.e., more matter than antimatter. In the GM it is assumed that during the formation of helium in the aftermath of the Big Bang that an equivalent surplus of antimatter was formed as neutrinos, which have $p = -1$, so that overall equal numbers of rishons and antirishons prevailed. This assumption is a consequence of the conservation of $p$ in all interactions.\\

Thus the ordinary matter present in the universe, prior to the fusion process into heavier elements, has essentially particle number $p = 0$. Since the additive quantum number $p$ is conserved in all interactions, this implies that the overall particle number of the universe will remain essentially as $p = 0$, i.e., symmetric in particle and antiparticle matter.\\ 

To summarize: the ordinary matter present in the universe has an overall particle number of $p = 0$, so that it contains equal numbers of both rishons and antirishons. This implies that the original antimatter created in the Big Bang is now contained within the stable composite leptons, i.e., electrons and neutrinos, and the stable composite quarks, i.e., the weak eigenstate up and down quarks, which comprise the protons and neutrons. The hydrogen, helium and heavier atoms all consist of electrons, protons and neutrons. This explains where all the antiparticles have gone. However, it does not explain why the universe consists primarily of hydrogen atoms and not antihydrogen atoms. It is suggested that the hydrogen-antihydrogen asymmetry may be understood as follows.\\

In the GM, antihydrogen consists of the same rishons and antirishons as does hydrogen, although the rishons and antirishons are differently arranged in the two systems. This implies that both hydrogen atoms and antihydrogen atoms should be formed during the aftermath of the Big Bang with about the same probability. In fact, estimates from the cosmic microwave background data suggest that for every billion hydrogen-antihydrogen pairs there was just one extra hydrogen atom. It is suggested that this extremely small difference, one extra hydrogen atom in $10^{9}$ hydrogen-antihydrogen pairs, may arise from statistical fluctuations associated with the complex many-body processes involved in the formation of either a hydrogen atom or an antihydrogen atom. The uniformity of the universe,$^{25}$ in particular, the lack of antihydrogen throughout the universe, indicates that the above statistical fluctuations took place prior to the `inflationary period'$^{26,27}$ associated with the Big Bang scenario.\\

\vspace{0.5cm}

\noindent {\bf 5. Conclusion and Discussion}\\

\noindent Within the framework of the GM of particle physics, it has been demonstrated that the matter-antimatter asymmetry problem may be understood in terms of the particle additive quantum number ($p$) and the composite nature of the leptons and quarks of the GM. The ordinary matter present in the universe has an overall particle number $p = 0$, so that it contains the same number of particles (rishons) as antiparticles (antirishons).\\

This implies that the original antimatter created in the Big Bang is now contained within the stable composite leptons, the electrons and neutrinos, and the stable composite quarks, the weak eigenstate up and down quarks that comprise the protons and neutrons. The hydrogen, helium and heavier atoms all consist of electrons, protons and neutrons.\\

Thus there is no matter-antimatter asymmetry in the present universe. However, there does exist a hydrogen-antihydrogen asymmetry: the present universe consists predominently of hydrogen atoms and virtually no antihydrogen atoms. In the SM this is tantamount to the matter-antimatter asymmetry, since both protons and electrons are assumed to be matter. In the GM this is not the case, since both hydrogen and antihydrogen atoms contain the same number of rishons as the number of antirishons.\\

An understanding of the hydrogen-antihydrogen asymmetry requires knowledge of the physical nature of the Big Bang but unfortunately this knowledge is currently far from complete. The prevailing model of the Big Bang is based upon the theory of general relativity:$^{28}$ extrapolation of the expansion of the universe backwards in time yields an infinite density and temperature at a finite time in the past (approximately 13.8 billion years ago). Thus the `birth' of the universe seems to be associated with a `singularity', which not only signals a breakdown of general relativity but also all the laws of physics. This leads to serious impediments to understanding the physical nature of the Big Bang and consequently the development of the hydrogen-antihydrogen asymmetry in the aftermath of the Big Bang.\\

In the SM, the singularity is generally ignored and it is assumed that the Big Bang initially produces numerous elementary particle-antiparticle pairs such as electron-positron pairs and quark-antiquark pairs by converting energy into mass according to $m = E/c^{2}$. Thus the early universe consisted of a soup of particle-antiparticle pairs continually being created and annihilated. Later, as the universe cooled following an inflationary period, the quarks and antiquarks would form protons, neutrons, antiprotons, antineutrons, etc., and eventually atoms of hydrogen, antihydrogen, helium and antihelium. These would later annihilate in pairs until only atoms of hydrogen and helium prevailed. In this scenario, it seems unlikely that either electrons or positrons would prevail so that neither hydrogen atoms nor antihydrogen atoms would prevail. This simply reflects that the creation and annihilation of electron-positron pairs constitute a unique process.\\

In the GM, neglecting the singularity, it is expected that the Big Bang would initially produce numerous elementary rishon-antirishon pairs. Then as the universe cooled following an inflationary period, the rishons and antirishons would form the leptons, quarks and their antiparticles and eventually atoms of hydrogen, antihydrogen, helium and antihelium. These would later annihilate in pairs until only atoms of hydrogen and helium prevailed. In this scenario, small statistical fluctuations occur at the rishon-antirishon level so that for every billion of hydrogen-antihydrogen pairs there is just one extra hydrogen atom. This simply reflects that the formation of electron-positron pairs is a more complex process in the GM than the unique process in the SM.\\

Thus there are two main conclusions: (1) there is no matter-antimatter asymmetry in the present universe and (2) it is suggested that the observed small hydrogen-antihydrogen asymmetry in the present universe can be understood in terms of statistical fluctuations associated with the complex many-body processes involved in the formation of either a hydrogen atom or an antihydrogen atom.\\

Finally it should be noted that if the Big Bang produced equal numbers of particles and antiparticles so that the initial state of the universe had particle number $p = 0$, then the GM {\it predicts} that the present state of the universe should also have $p = 0$, since particle number $p$ is conserved in all interactions.\\

\vspace{0.5cm}

\noindent {\bf Acknowledgments}\\
 
\noindent I am grateful to N. H. Fletcher, D. Robson, J. M. Robson and L. J. Tassie for helpful discussions.

\vspace{0.5cm}

\noindent {\bf References}

\noindent \hspace*{0.2cm}1. P. A. R. Ade et al., (Planck Collaboration) arXiv:1502.01590v1 [astro-p
\newline
\hspace*{0.8cm}h.CO] 25 Feb 2015 to be published in {\it Astronomy and Astrophysics}.
\newline
\hspace*{0.2cm}2. B. A. Robson, {\it Int. J. Mod. Phys. E} {\bf 22} (2013) 1350067.
\newline
\hspace*{0.2cm}3. B. A. Robson, {\it Int. J. Mod. Phys. E} {\bf 24} (2015) 1550012.
\newline
\hspace*{0.2cm}4. B. A. Robson, The Generation Model of Particle Physics in {\it Particle Physics}, 
\newline
\hspace*{0.8cm}Ed. E. Kennedy (InTech Open Access Publisher, Rijeka, Croatia, 2012).
\newline
\hspace*{0.2cm}5. K. Gottfried and V. F. Weisskopf, {\it Concepts of Particle Physics} Vol. 1 Oxford
\newline
\hspace*{0.8cm}University Press, New York, 1984).
\newline
\hspace*{0.2cm}6. D. J. Gross and F. Wilczek, {\it Phys. Rev. Lett.} {\bf 30} (1973) 1343.
\newline
\hspace*{0.2cm}7. H. D. Politzer, {\it Phys. Rev. Lett.} {\bf 30} (1973) 1346.
\newline
\hspace*{0.2cm}8. F. Wilczek, {\it Nature} {\bf 433} (2005) 239. 
\newline
\hspace*{0.2cm}9. R. Morris, {\it Cosmic Questions}, (John Wiley and Sons, New York 1993).
\newline
10. A. D. Sakharov, {\it JETP} {\bf 5} (1967) 24.
\newline
11. G. R. Farrar and M. E. Shaposhnikov, {\it Phys. Rev. Lett.} {\bf 70} (1993) 2833.
\newline
12. J. H. Christenson, J. W. Cronin, V. I. Fitch and R. Turlay, {\it Phys. Rev. Lett.}
\newline
\hspace*{0.8cm}{\bf 13} (1964) 138.
\newline
13. N. Cabibbo, {\it Phys. Rev. Lett.} {\bf 10} (1963) 531.
\newline
14. M. Kobayashi and T. Maskawa, {\it Prog. Theor. Phys.} {\bf 49} (1973) 652.
\newline
15. M. Gell-Mann, {\it Phys. Lett.} {\bf 8} (1964) 214.
\newline
16. G. Zweig, {\it CERN reprts} {\bf 8182/TH 401} and {\bf 8419/TH 412} (1964).
\newline
17. M. Dine and A. Kusenko, {\it Rev. Mod. Phys.} {\bf 76} (2004) 1.
\newline
18. B. A. Robson, {\it Int. J. Mod. Phys. E} {\bf 11} (2002) 555.
\newline
19. B. A. Robson, {\it Int. J. Mod. Phys. E} {\bf 13} (2004) 999.
\newline
20. B. A. Robson, {\it Int. J. Mod. Phys. E} {\bf 14} (2005) 1151.
\newline
21. P. W. Evans and B. A. Robson, {\it Int. J. Mod. Phys. E} {\bf 15} (2006) 617.
\newline
22. B. A. Robson, {\it Int. J. Mod. Phys. E} {\bf 20} (2011) 733.
\newline
23. B. A. Robson, {\it Advances in High Energy Physics} {\bf 2013} (2013) 341738.
\newline
24. B. A. Robson, {\it Int. J. Mod. Phys. E} {\bf 18} (2009) 1773. 
\newline
25. D. Lincoln, {\it Understanding the Universe from Quarks to the Cosmos}, rev. edn. 
\newline
\hspace*{0.8cm}(World Scientific, Singapore, 2012).
\newline
26. A. H. Guth, {\it Phys. Rev. D} {\bf 23} (1981) 347.
\newline
27. A. D. Linde, {\it Phys. Lett. B} {\bf 108} (1982) 389.
\newline
28. A. Einstein, {\it Ann. Physik} {\bf 49} (1916) 769.
   
\end{document}